\documentclass[modern,tighten]{aastex62}

\newcommand{\nicer}{\textit{NICER}{}}
\newcommand{\ngc}{NGC 300 ULX-1{}}

\graphicspath{{./}{figures/}}

\shorttitle{Anti-glitches in NGC 300 ULX-1}
\shortauthors{Ray et al.}

\begin{document}

\title{Anti-glitches in the Ultraluminous Accreting Pulsar \ngc{} \\Observed with \textit{NICER}}

\correspondingauthor{Paul S. Ray}
\email{paul.ray@nrl.navy.mil}

\author[0000-0002-5297-5278]{Paul S. Ray}
\affil{Space Science Division, U.S. Naval Research Laboratory, Washington, DC 20375, USA}
\author[0000-0002-6449-106X]{Sebastien Guillot}
\affil{IRAP, CNRS, 9 avenue du Colonel Roche, BP 44346, F-31028 Toulouse Cedex 4, France}
\affil{Universit\'{e} de Toulouse, CNES, UPS-OMP, F-31028 Toulouse, France}
\author[0000-0002-6089-6836]{Wynn C.G. Ho}
\affil{Department of Physics and Astronomy, Haverford College, 370 Lancaster Avenue, Haverford, PA 19041, USA}
\affil{Mathematical Sciences, Physics and Astronomy, and STAG Research Centre, University of Southampton, Southampton SO17 1BJ, UK}
\author{Matthew Kerr}
\affil{Space Science Division, U.S. Naval Research Laboratory, Washington, DC 20375, USA}
\author[0000-0003-1244-3100]{Teruaki Enoto}
\affil{The Hakubi Center for Advanced Research and Department of Astronomy, Kyoto University, Kyoto 606-8302, Japan}
\author{Keith C.~Gendreau}
\affil{X-Ray Astrophysics Laboratory, NASA Goddard Space Flight Center, Greenbelt, MD 20771, USA}
\author{Zaven Arzoumanian}
\affil{X-Ray Astrophysics Laboratory, NASA Goddard Space Flight Center, Greenbelt, MD 20771, USA}
\author[0000-0002-3422-0074]{Diego Altamirano}
\affil{Physics and Astronomy, University of Southampton, Southampton, Hampshire SO17 1BJ, UK}
\author[0000-0002-9870-2742]{Slavko Bogdanov}
\affil{Columbia Astrophysics Laboratory, Columbia University, 550 West 120th Street, New York, NY 10027, USA}
\author{Robert Campion}
\affil{Chesapeake Aerospace, Grasonville, MD 21638, USA}
\author{Deepto Chakrabarty}
\affil{MIT Kavli Institute for Astrophysics and Space Research, Massachusetts Institute of Technology, Cambridge, MA 02139, USA}
\author{Julia S. Deneva}
\affil{George Mason University, resident at the Naval Research Laboratory, Washington, DC 20375, USA}
\author[0000-0002-6789-2723]{Gaurava K. Jaisawal}
\affil{National Space Institute, Technical University of Denmark, 
  Elektrovej 327-328, DK-2800 Lyngby, Denmark}
\author{Robert Kozon}
\affil{Code 681, NASA Goddard Space Flight Center, Greenbelt, MD 20771, USA}
\author{Christian Malacaria}
\affiliation{NASA Marshall Space Flight Center, NSSTC, 320 Sparkman Drive, Huntsville, AL 35805, USA}\thanks{NASA Postdoctoral Fellow}
\affiliation{Universities Space Research Association, NSSTC, 320 Sparkman Drive, Huntsville, AL 35805, USA}
\author[0000-0001-7681-5845]{Tod E. Strohmayer}
\affil{Astrophysics Science Division and Joint Space-Science Institute, NASA Goddard Space Flight Center, Greenbelt, MD 20771, USA}
\author[0000-0002-4013-5650]{Michael T. Wolff}
\affil{Space Science Division, U.S. Naval Research Laboratory, Washington, DC 20375, USA}



\begin{abstract}

We {present evidence for} three spin-down glitches (or `anti-glitches') in the ultraluminous accreting X-ray pulsar \ngc\ in timing observations made with the \textit{Neutron Star Interior Composition Explorer} (\textit{NICER}).  {Our timing analysis reveals} three sudden spin-down events of magnitudes $\Delta\nu = -23, -30,$ and $-43 \,\mu$Hz (fractional amplitudes $\Delta\nu/\nu = -4.4, -5.5,$ and $-7.7 \times 10^{-4}$). {We determined fully phase-coherent timing solutions through the first two glitches, 
giving us high confidence in their detection, while the third candidate glitch is somewhat less secure.}
These are larger in magnitude (and opposite in sign) than any {known} radio pulsar glitch. 
This may be caused by the prolonged rapid spin-up of the pulsar causing a sudden transfer of angular momentum between the superfluid and non-superfluid components of the star.  We find no evidence for profile or spectral changes at the epochs of the glitches, supporting the conclusion that these
are due to the same process as in normal pulsar glitches, but in reverse.
\end{abstract}



\section{Introduction} \label{sec:intro}
Ultraluminous X-ray sources (ULXs) are extragalactic, bright X-ray point sources with super-Eddington luminosities for isotropic emission due to accretion onto a  neutron star or stellar-mass black hole \citep[see][for a recent review]{2017ARA&A..55..303K}. Following the discovery of X-ray pulsations from a handful of ULXs with periods ranging from 0.42 sec to $\sim$31 sec \citep{bachetti14,fuerst16,israel17a,israel17b}, it has been suggested that the central engines of most ULXs could be highly-magnetized accreting neutron stars, based on theoretical \citep{king16,mushtukov17} or observational arguments \citep{koliopanos17,walton18a}. Further evidence comes from the claimed discovery of Cyclotron Resonant Scattering Features (CRSF) in two ULXs \citep{brightman18,walton18b}, albeit one of them with no detected pulsations.

The ULX in the nearby spiral galaxy NGC 300 ($D=1.87\pm0.12$ Mpc; \citealt{rizzi06}) was mistakenly classified as a supernova (SN 2010da) upon its discovery \citep{2010CBET.2289....1M}. Follow-up \textit{Chandra} observations showed that the system is instead a high-mass X-ray binary (HMXB) undergoing an outburst \citep{2011ApJ...739L..51B}. Using observations obtained in 2016 December with \textit{XMM-Newton} and \textit{NuSTAR}, \citet{carpano18} discovered X-ray pulsations at a frequency of $\nu=31.6$ mHz ($P = 31.6$ s, first reported in \citealt{ATel11158}), thus revealing the compact object as a neutron star. They also found that the pulsar is spinning up very rapidly with $\dot{\nu}=5.57\times10^{-10}$ Hz s$^{-1}$.

This article reports on the regular monitoring of \ngc{} with the \emph{Neutron Star Interior Composition Explorer} (\nicer, \citealt{2016SPIEGendreau}), and the discovery of three apparent spin-down glitches that occurred over a span of about 120 days.  In the many hundreds of radio, X-ray, and $\gamma$-ray pulsars monitored for glitches, all of the hundreds of detected glitches are spin-up glitches ($\Delta\nu>0$; see, e.g., \citealt{2011MNRAS.414.1679E,2013MNRAS.429..688Y,2017A&A...608A.131F}). Among magnetars, observations reveal approximately twenty glitches, consisting of both spin-up and spin-down ones {(see, e.g., \citealt{2014ApJ...784...37D})}.  The largest magnetar spin-down glitch (with $\Delta\nu=-21\,\mbox{$\mu$Hz}$) is inferred to possibly have occurred during a giant flare of SGR~1900+14 \citep{1999ApJ...524L..55W}, while the second largest was over a hundred times smaller ($\Delta\nu\sim -0.09\,\mbox{$\mu$Hz}$) and accompanied by radiative changes \citep{Archibaldetal2013}. The glitches described above all occur in non-accreting pulsars, whose long-term spin frequencies decrease with time ($\dot{\nu}<0$) due to rotational energy loss from electromagnetic dipole radiation \citep{1968Natur.219..145P,1969Natur.221..454G}. For pulsars that accrete matter from a binary companion, the torque due to accretion can cause the neutron star to spin-up or spin-down and is usually much stronger than the dipole radiation torque. There are only a few reported glitches in accreting pulsars: two spin-up glitches with $\Delta\nu\approx 1\,\mbox{$\mu$Hz}$, each occurring during an accretion outburst episode \citep{2004ApJ...613.1164G,2017MNRAS.471.4982S}, and possibly two spin-down glitches \citep{2009A&A...506.1261K}. 




\section{Observations}
\label{sec:obs}
In response to a series of Astronomer's Telegrams reporting the discovery of pulsations and the period  evolution of \ngc\ 
\citep{ATel11158,ATel11174,ATel11179,ATel11228,ATel11229,ATel11282,ATel11285}, \nicer{} observed \ngc{} on 2018 February 6 for 1.6 ks.  Following that, the source entered \nicer's solar exclusion zone and became unobservable until 2018 May 1, when \ngc{} was again sufficiently far from the Sun. \nicer{} then started a monitoring campaign on the source, primarily designed to discover evidence of an orbital period. \nicer{} generally has a visibility window of a few hundred to $\sim$2000 seconds in one segment of an orbit and observations were generally made during 1--10 consecutive orbital segments per day (most often 2, 3, or 4).  Here we present an analysis of \nicer{} ObsIDs 1034200101 (2018 February 6) through 1034200198 (2018 October 20).

The X-ray Timing Instrument (XTI) of \nicer{} provides $\approx$1900 cm$^{2}$ of collecting area at 1.5 keV thanks to its 56 co-aligned X-ray concentrating optics paired with single-pixel silicon drift detectors (SDD) working in the 0.2--12 keV energy band (52 detectors are operational on orbit). The GPS time tagging of photons permits obtaining an accuracy better than 100 ns root mean square (RMS) \citep{2016SPIELaMarr,2016SPIEPrigozhin}. 

The data were processed following standard procedures using the \nicer{} Data Analysis Software 
{\tt NICERDAS} version 4, together with {\tt HEASOFT} version 6.24. The standard filtering criteria 
define the good time intervals for which the \nicer\ pointing offset is $<0.015\deg$ from the nominal source position, 
the source is at least $30\deg$ from the Earth limb (at least $40\deg$ in the case of a Sun-illuminated Earth), 
and intervals when \nicer\ was not within the boundaries of the South Atlantic Anomaly. 
In a few observations, we masked ``hot'' detectors exhibiting out-of-family count rates. Also, for ObsID 1034200101, we had to remove MPU3 \footnote{MPU: Measurement Power Unit; each MPU connects 8 of the Focal Plane Modules, each containing one SDD} because of extreme count rates caused by solar loading on the detectors, and relaxed the filtering for the bright Earth to $30\deg$.  For each observation, we selected events in the 0.4--12 keV band and barycentered the data using the DE405 Solar System ephemeris from JPL and the coordinates of the ULX from \textit{Chandra} imaging data (R.A. = 13.770208\degr, Decl. = $-37.695417$\degr, \citealt{2011ApJ...739L..51B}). All barycentric times and frequencies in this work use the Barycentric Dynamical Time (TDB) time scale. Some of the observations exhibited background flares when the International Space Station was at extreme northern or southern latitudes. We applied a count rate cut of 7.5 c/s on non-overlapping 8-second intervals to exclude the times of such flares. 

\begin{figure}
\centering
\includegraphics[width=4.0in]{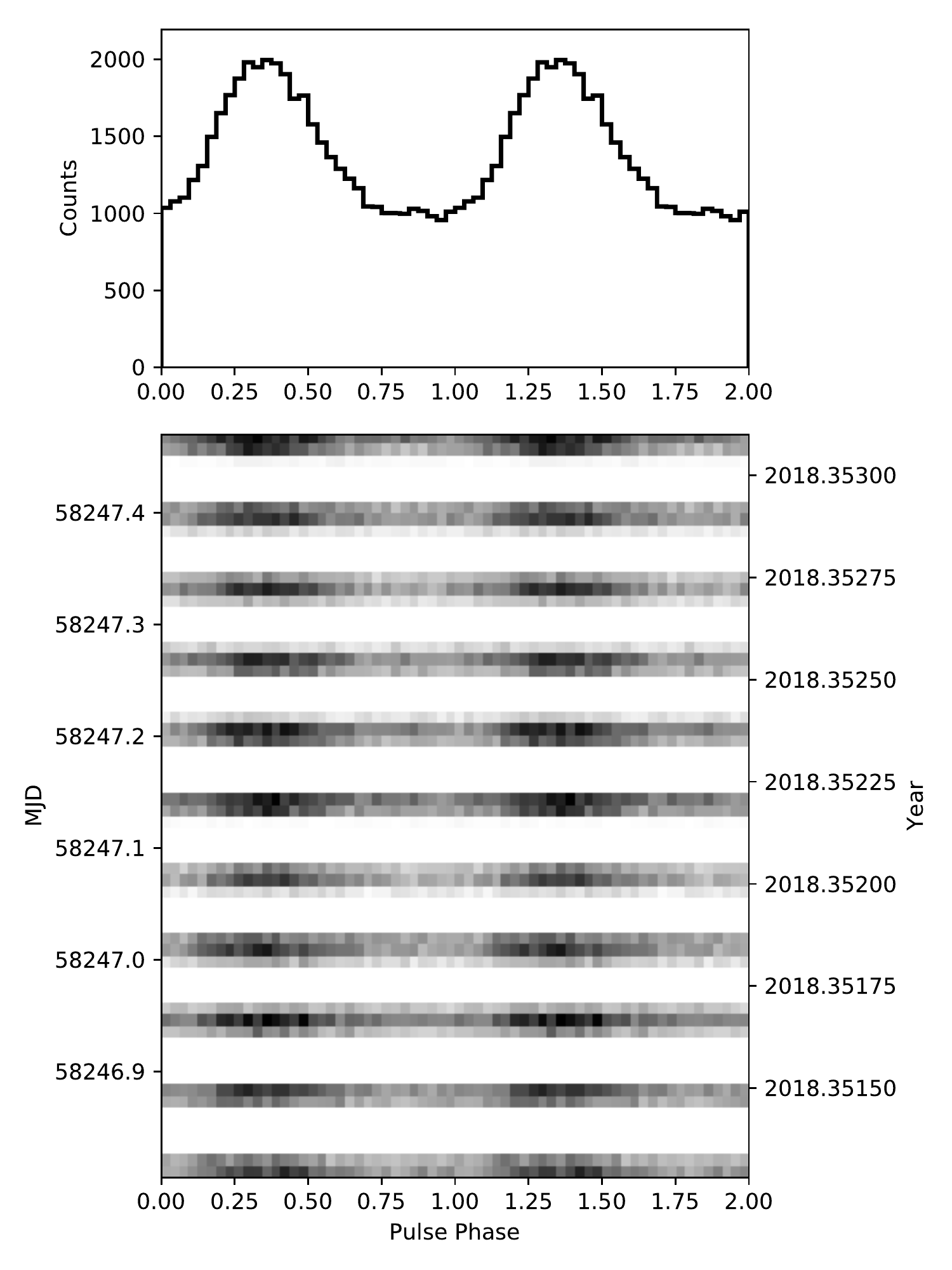}
.\caption{Phaseogram of \ngc{} in the 0.4--12 keV band, from the \nicer{} observations around MJD 58246. \label{fig:profile}}
\end{figure}

\section{Timing}
\label{sec:timing}

{The timing analysis demonstrates the power of NICER for timing studies of faint X-ray pulsars. NICER's flexible scheduling enabled monitoring of the source in a way that maximized the ability to do precise timing. We scheduled observations to cover two or more consecutive ISS orbits, spreading a relatively small amount of observing time over a duration long enough to make a precise frequency measurement, but not so long that the window function caused significant ambiguity in the pulse counting between data segments.  These precise frequency measurements allowed us to bootstrap a fully-coherent timing solution. In the following, we present the timing analysis of the pulsations, working from coarse frequency measurements to phase-coherent timing.}

For the initial timing analysis, we binned the data at 1/128 seconds and used the PRESTO software package \citep{PRESTO} to perform a periodicity search over a range of possible accelerations. In essentially every observation until 2018 August 22, when the source seems to have faded significantly, the ULX pulsations (at around 53--57 mHz) were detected as the highest significance peak coming out of the search. This allowed us to confirm that the period evolution was continuing as previously reported.
In addition, we confirmed the broad Gaussian pulse profile of this ULX reported previously \citep{carpano18}.
Figure~\ref{fig:profile} shows the pulse profile from MJD 58246, \nicer's densest day of observations. The profile is reasonably well modeled as a Gaussian with full width at half maximum (FWHM) of 0.35 cycles.

To go from initial periodicity detections to precise period measurements, we employed a maximum likelihood technique to measure pulsed frequencies and significance. In this case, given a selection of photons from a segment of the observation, we searched over a grid of trial $\nu$ values. For each trial we computed the pulse phase of each photon according to a model with the trial $\nu$ (and $\dot{\nu}$ 
fixed at the long-term average, $4.16\times 10^{-10}$ Hz s$^{-1}$). We model the probability distribution of this set of phases using a Gaussian template of FWHM 0.35 of the pulse period and subsequently maximized the likelihood to determine the optimal values of the Gaussian peak position and amplitude.  By profiling 
over these nuisance parameters, we arrive at a profile likelihood for $\nu$; see \citet{2015ApJ...814..128K} and \citet{r+11} for a discussion of similar likelihood techniques. This finally yielded a maximum likelihood estimate for the pulse frequency at that observation, as well as an estimate of the measurement uncertainty determined by the frequency range over which the log likelihood decreased by 0.5.

The rapid and noisy period evolution, combined with the window function of the observations, made precision timing difficult, so we developed a bootstrapping technique to make precise pulse frequency measurements.  The first step was making likelihood 
frequency measurements for each nearly-contiguous segment of \nicer{} data (i.e., with no gaps greater than 1000 s and a total span less than 3000 s). These measurements are low-precision, but unambiguous. 
We note that the last significant pulsed detection was at MJD 58352.5 (2018 August 22; ObsID 1034200183). All subsequent observations show no strong evidence of pulsations from \ngc{}. This includes an 800 s observation at MJD 58351.5, followed by a gap of a month, then 21 observations between MJD 58386 and 58411\footnote{{While this paper was in review, monitoring continued and the source re-brightened allowing some additional pulsation detections after MJD 58450. These were not sufficient to establish a coherent timing solution or to search for additional glitches. For an analysis of those data and the implications of the spin evolution, see \citet{vas}. }}.

\begin{figure}
\includegraphics[width=6.5in]{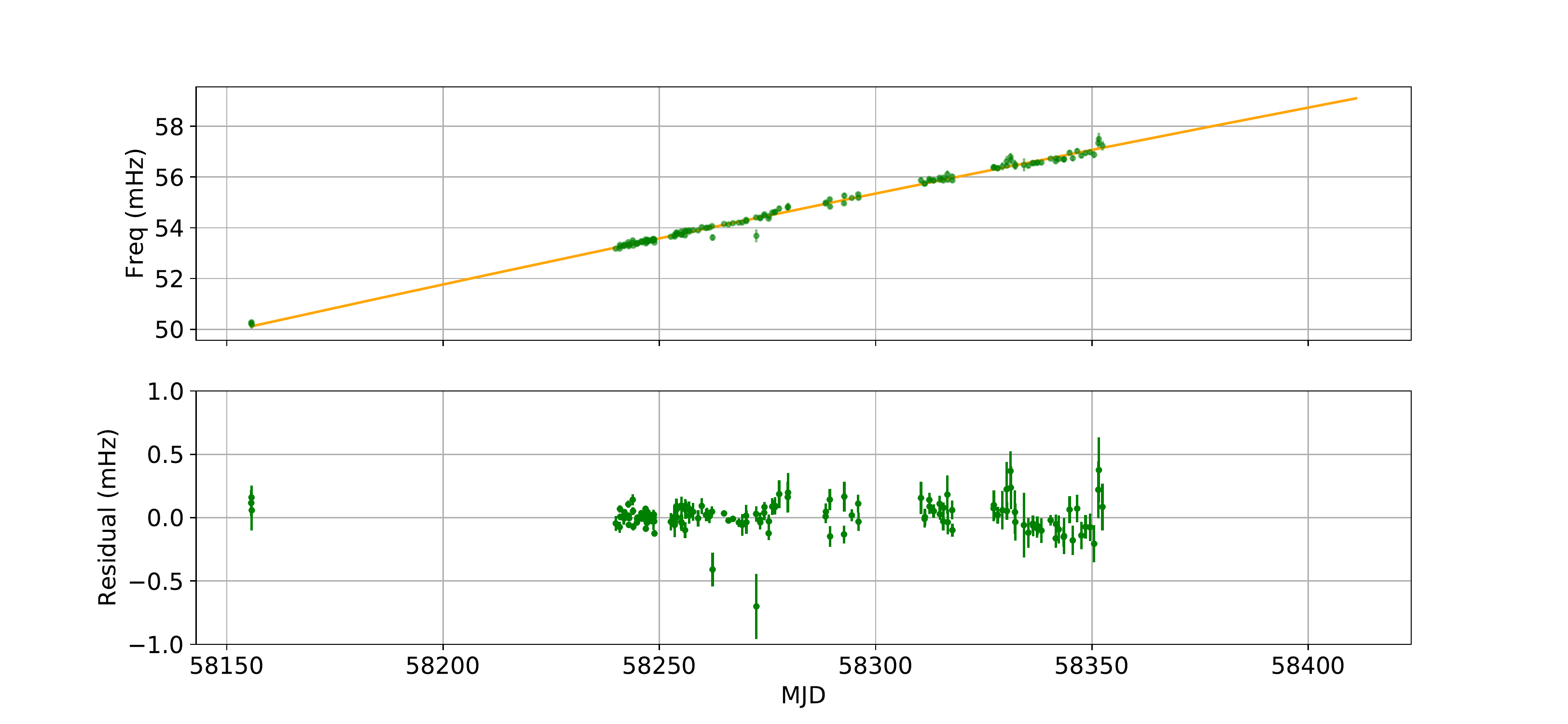}
\caption{Pulse frequency evolution of \ngc\ as measured with \nicer{} from each 
short, nearly-contiguous data segment. The line is the fitted model as described in the text, and the bottom panel shows the residuals to the fit.  
Note that glitches are not readily apparent here because of the poor precision of the 
individual period measurements. 
\label{fig:shortfreqs}}
\end{figure}

\begin{figure}
\includegraphics[width=6.5in]{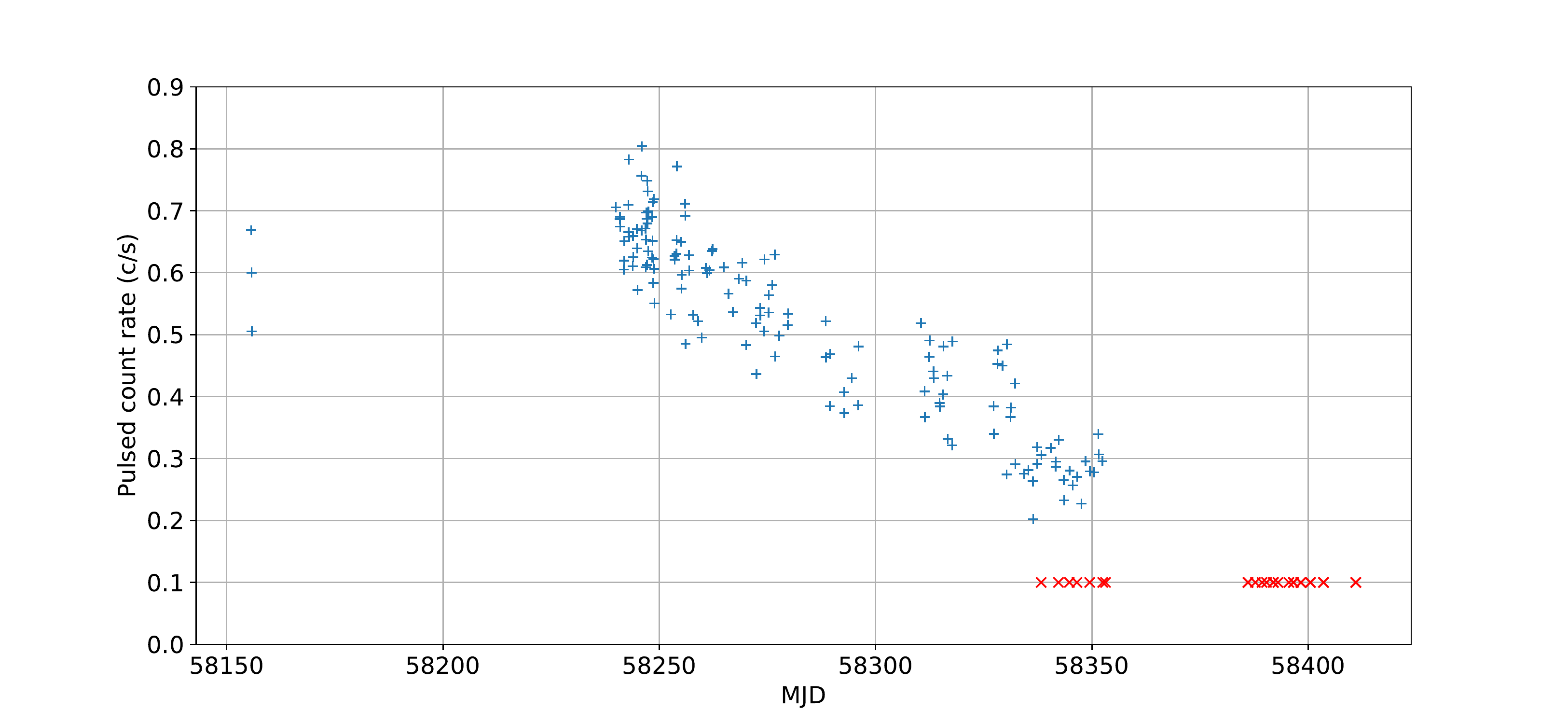}
\caption{Pulsed count rate for each \nicer{} observation where pulsations from \ngc\ were significantly detected. The count rates near MJD 58155 have
been scaled by 7/6 to account for the exclusion of MPU3 in the analysis. Observations with at least 500 s of exposure where no significant pulsation was detected are marked with red $\times$ symbols.\label{fig:rate}}
\end{figure}

The measured frequencies for each segment of \nicer{} data are shown in Figure \ref{fig:shortfreqs}.
Fitting the pulse frequency evolution to a simple model with two frequency derivatives 
gives the best fit $\nu = \nu_0 + \dot{\nu}(t-t_0) + \frac{1}{2}\ddot{\nu}(t-t_0)^{2}$, 
where $\nu_0 = 53.350(5)$ mHz, $\dot{\nu} = 4.16(5) \times 10^{-10}$ Hz s$^{-1}$, 
$\ddot{\nu} = -2.5(12) \times 10^{-18}$ Hz s$^{-2}$ and
$t_0$ = MJD(TDB) 58243.515. Here and elsewhere in the paper the numbers in parentheses are the {1$\sigma$} uncertainty in the last digit.  If the spindown proceeds as $\dot{\Omega} \propto \Omega^n$ (where $\Omega$ is the angular velocity), the `braking index' $n$ is defined as $n =\nu\ddot{\nu}/\dot{\nu}^2$. For \ngc{}, $n$ is $-0.8(4)$. The small magnitude of $n$ indicates that we are likely seeing the secular spinup of the system and the $\ddot{\nu}$ is not strongly contaminated by timing noise or changing Doppler shift from an orbit. An analysis of the long-term spinup in terms of accretion torque theory is presented in \citet{Georgios}. 
The fitting procedure also yields a pulsed count rate for each observation; these are shown in Figure \ref{fig:rate}.



To make more precise frequency measurements, we repeated the likelihood analysis above, with 
expanded observation segments with a maximum span of 15000 s (limiting the duration to 3 \nicer{} orbits). In this stage, there are potential pulse counting ambiguities due to the long gaps between orbits in some observations. However, in nearly all cases, measurements that suffered from an off-by-one counting error between observation subsets were very clearly 
identified from the short-segment likelihood frequency estimates and 
the known position of the sidelobes from the observation window function.

\begin{figure}
\centering
\includegraphics[width=6.5in]{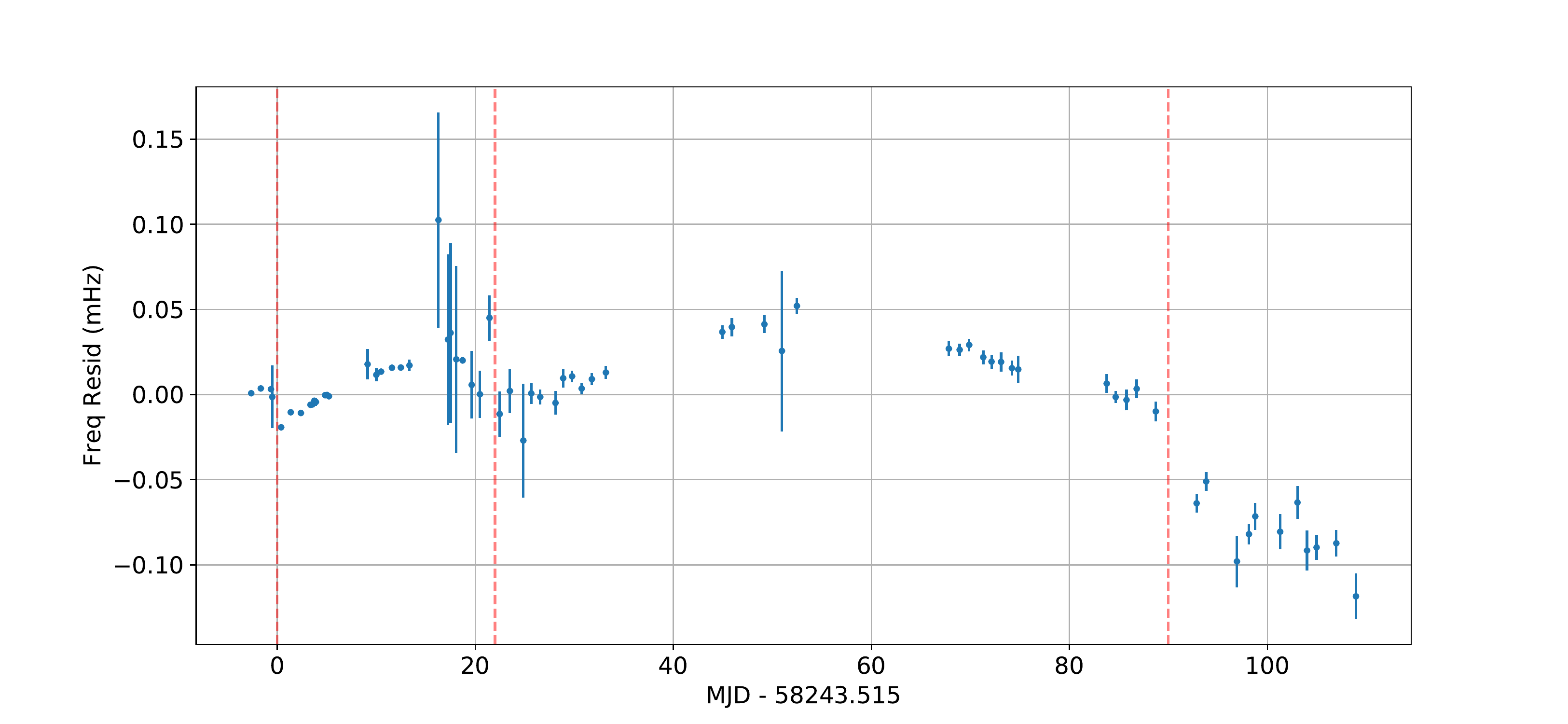}
\caption{\nicer{} pulse frequency measurements for \ngc{}, after subtraction of a simple model with a constant frequency derivative. The candidate glitch epochs are marked with red dashed lines.  \label{fig:plotfit1}}
\end{figure}

These precise frequency measurements, excluding the single observation in 2018 February, are shown in Figure \ref{fig:plotfit1}, where a constant frequency derivative has been subtracted to allow the fine structure to be seen.  This reveals relatively smooth frequency evolution which shows some evidence for several negative frequency steps, which we identify as candidate glitches. However, the sampling and large uncertainties in some of the measurements, coupled with the rapid underlying spin evolution, makes this result tentative, requiring confirmation from the phase-coherent analysis presented in \S\ref{sec:coherent}.

\begin{figure}
\centering
\includegraphics[width=6.5in]{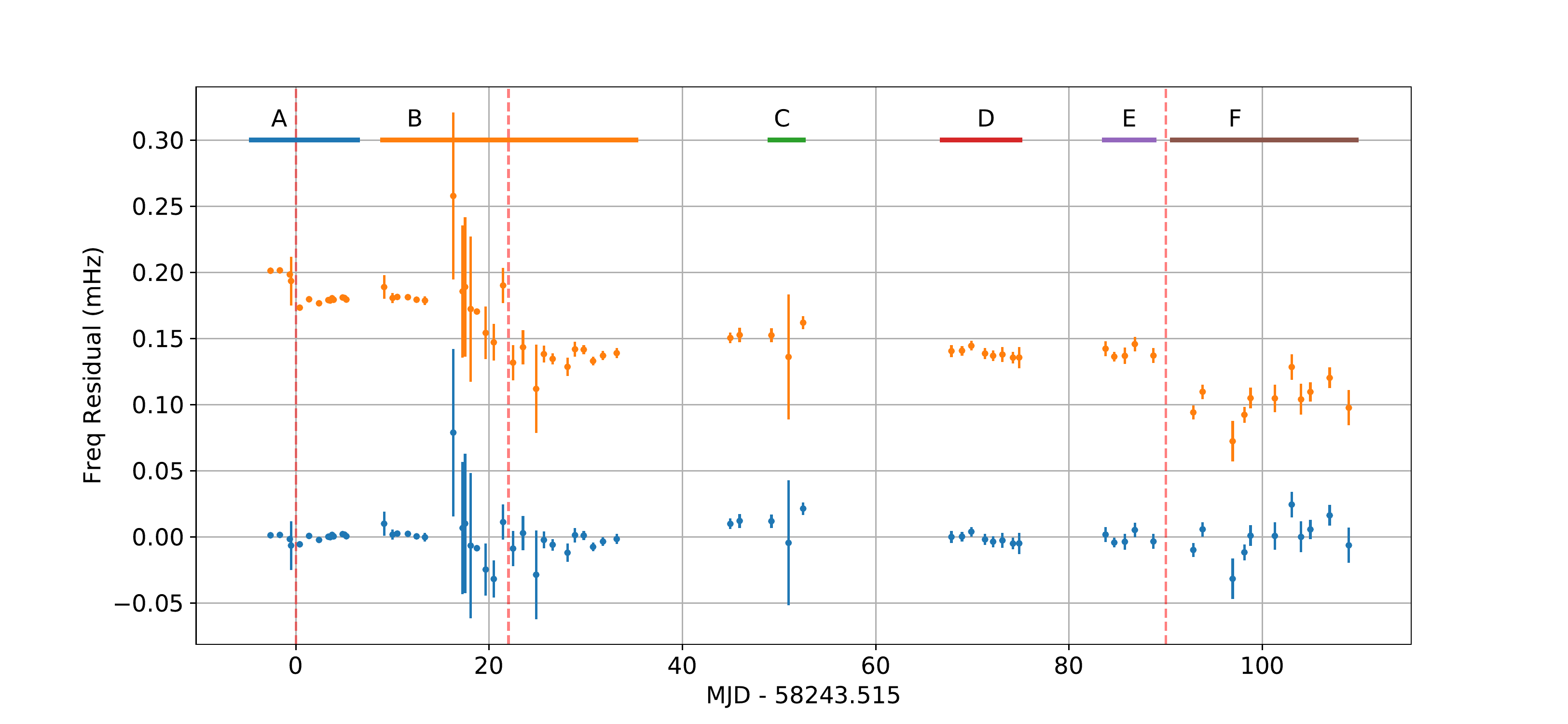}
\caption{Residuals to best-fit 3-glitch model for the \ngc\ frequency evolution. The upper (orange) points are the same data (offset by $2.0 \times 10^{-4}$ Hz) with the glitch magnitudes set to 0, to show the location and amplitude of the glitches.  {For reference, the time ranges segments A--F used in the coherent analysis in \S\ref{sec:coherent} are also shown.}\label{fig:plotfit4}}
\end{figure}

We attempted to fit the frequency evolution with a secular evolution up through $\ddot{\nu}$ and three glitches. The residuals to this fit are shown in Figure \ref{fig:plotfit4}.  There are clearly some unmodeled residuals, particularly in the high-precision measurements between glitches 1 and 2, and glitches 2 and 3. This could potentially be modeled as frequency derivative steps at the glitch, or exponentially-decaying glitch `recoveries'. Both of these phenomena are observed in rotation-powered pulsar glitches, but it is not clear which applies here, so we keep the model as simple as possible. The model parameters are shown in Table \ref{tab:par}.  We note that these model parameters yield a braking index $n=-0.9(2)$.

\begin{deluxetable}{lr}
\tablecaption{Incoherent timing parameters\label{tab:par}}
\tablehead{
\colhead{Parameter} & \colhead{Value} \\
}
\startdata
Epoch (MJD) & 58243.515 \\
$\nu$ (mHz) &  53.3533(7) \\
$\dot{\nu}$ (Hz s$^{-1}$) & $4.40(1)\times 10^{-10}$ \\
$\ddot{\nu}$ (Hz s$^{-2}$) & $-3.4(9) \times 10^{-18}$ \\
${\nu}^{(3)}$ (Hz s$^{-3}$) & $-1.1(3) \times 10^{-24}$ \\
Glitch 1 Epoch (MJD) & 58243.515 \\
Glitch 1 $\Delta\nu$ (Hz) & $-2.1(1) \times 10^{-5}$ \\
Glitch 2 Epoch (MJD) & 58265.5 \\
Glitch 2 $\Delta\nu$ (Hz) & $-3.8(2) \times 10^{-5}$ \\
Glitch 3 Epoch (MJD) & 58333.5 \\
Glitch 3 $\Delta\nu$ (Hz) & $-3.7(5) \times 10^{-5}$ \\
\enddata
\end{deluxetable}

\subsection{Search for profile variations}

\begin{figure}
\centering
\includegraphics[width=6.5in]{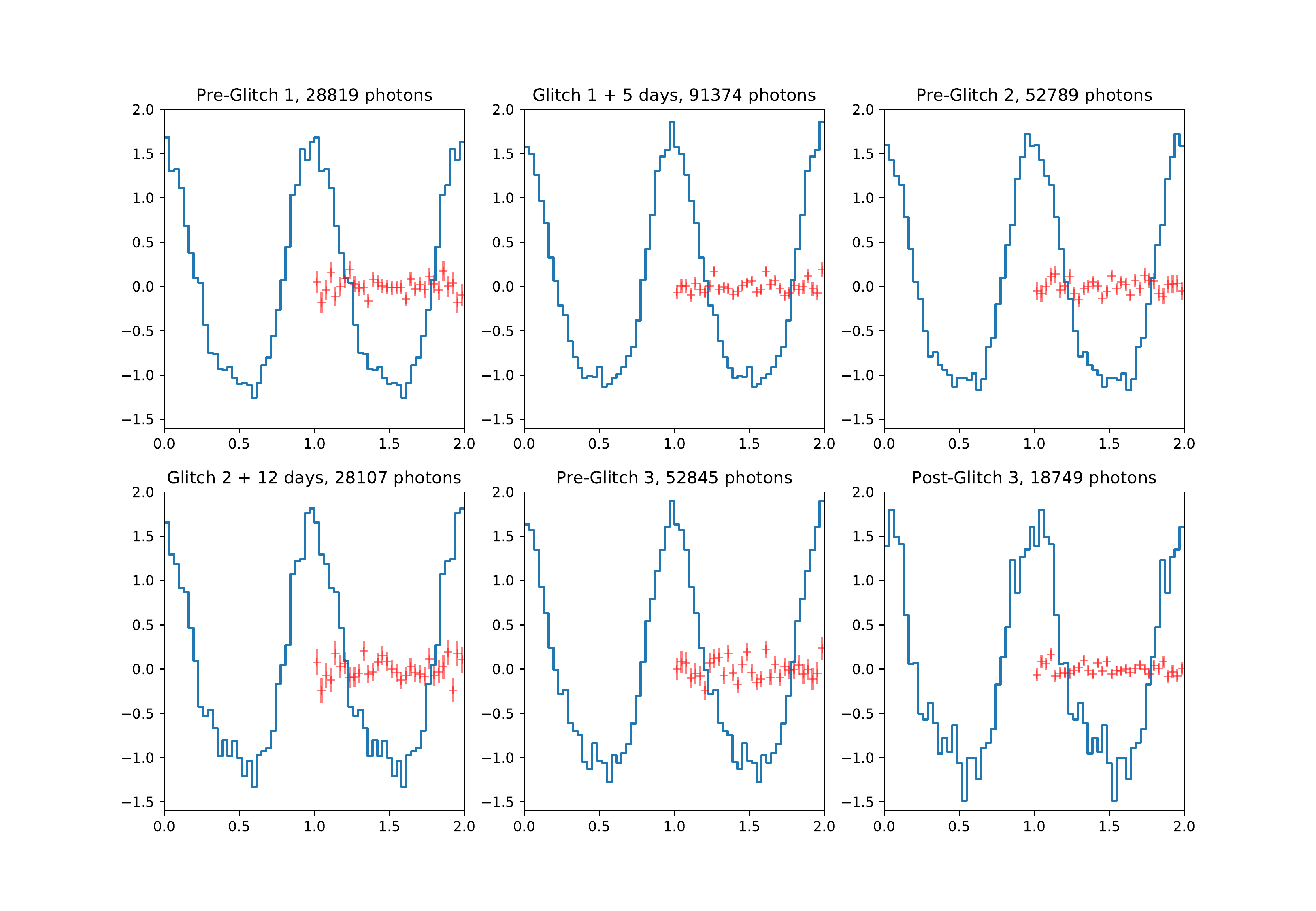}
\caption{Pulse profiles (0.4--12 keV) from 6 different sections of data. Each profile has the mean subtracted and is normalized by the standard deviation.  Two cycles are shown. The red points with error bars on the second cycle are the residuals of each profile to a model profile fitted to all of the data.\label{fig:profiles}}
\end{figure}

In most radio pulsar glitches the pulse profile does not change at the epoch of the glitch, 
{though} significant profile changes are observed at the epoch of some magnetar glitches (e.g., see \citealt{Archibaldetal2013}) {and radiative changes were seen after a glitch in one high magnetic field pulsar PSR J1119$-$6127 \citep{wje11}}. 
To help constrain the models for the \ngc{} glitches, we looked for profile variations.
To do this in a timing-model-independent way, we considered observations where the pulsations were 
strongly detected (6$\sigma$ or more). For each segment, we fitted a single gaussian template to determine 
phase 0 and generated a set of phases using that as a reference.  We then collected all of these 
phases from 6 different 
sections of the \nicer{} monitoring observation, splitting the data at the glitches, and at the first
significant data gap after each glitch.  We binned these phases into 40-bin histograms, subtracted
the mean and divided by the standard deviation, to account for the changing background and evolving 
pulsed count rate.  These 6 pulse profiles are displayed in Figure \ref{fig:profiles}. We also fitted a 3-gaussian template model to the complete dataset and display the residuals of each profile to that mean shape. We 
see no signficant pulse profile changes during our observations. 

\section{Coherent Timing Analysis}
\label{sec:coherent}

Using the observed frequency evolution, we computed pulse times of arrival (TOAs) for each 150 s segment of \nicer{} data (300 s for segment F, as defined below, where the count rate was significantly lower), using the unbinned maximum likelihood technique described in \cite{r+11}.  The reference phase for each TOA is the peak of the gaussian pulse. Typical TOA uncertainties are 0.5 seconds, or about 3\% of a pulse period.

\begin{figure}
\includegraphics[width=3.1in]{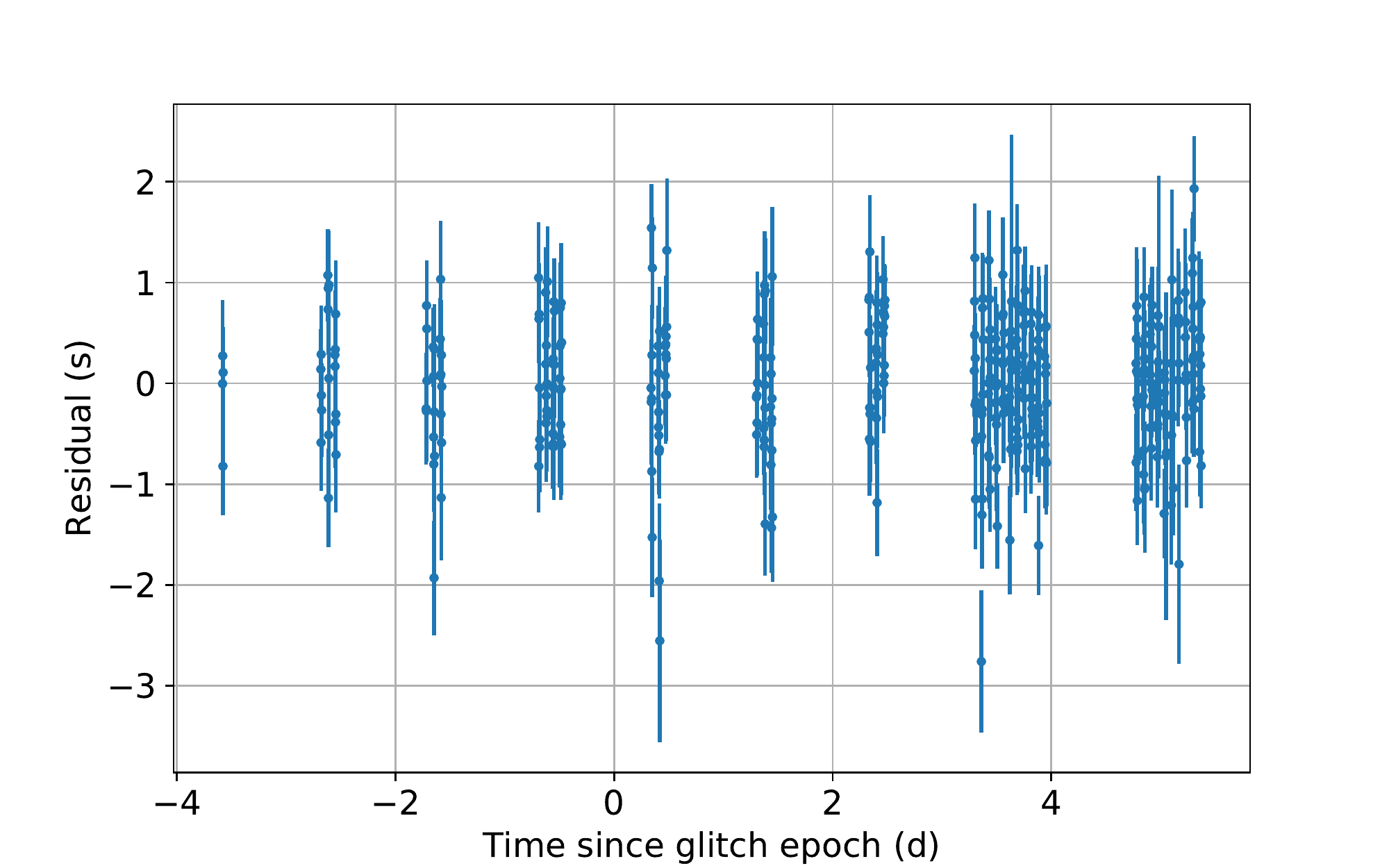}
\includegraphics[width=3.1in]{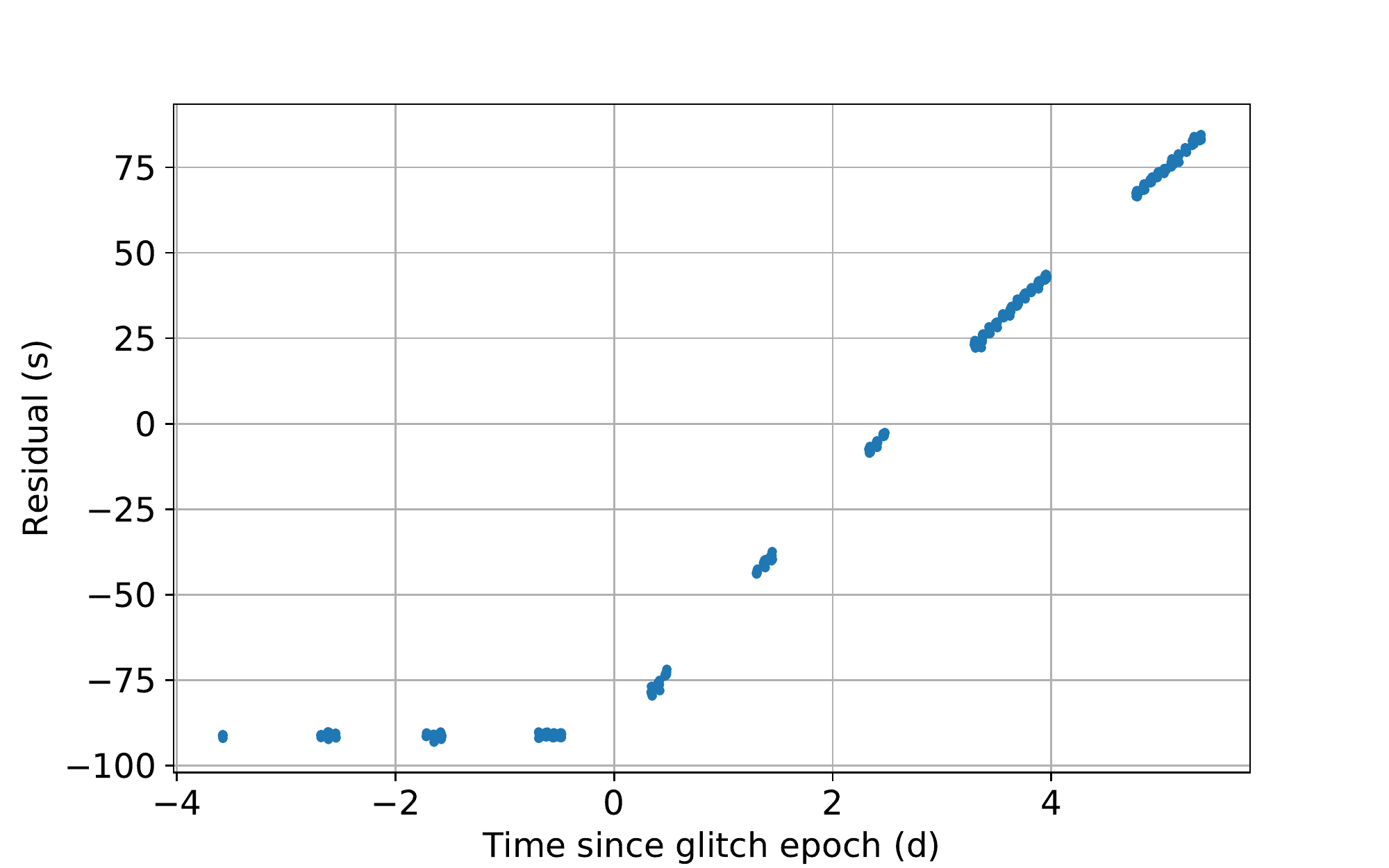}
\caption{Coherent timing of glitch 1 (segment A). Left: Residuals to a timing model including the glitch. Right: Residuals to the same model with $\Delta\nu$, $\Delta\dot{\nu}$ set to 0 to show the effect of the glitch.\label{fig:resids}}
\end{figure}

\begin{figure}
\includegraphics[width=3.0in]{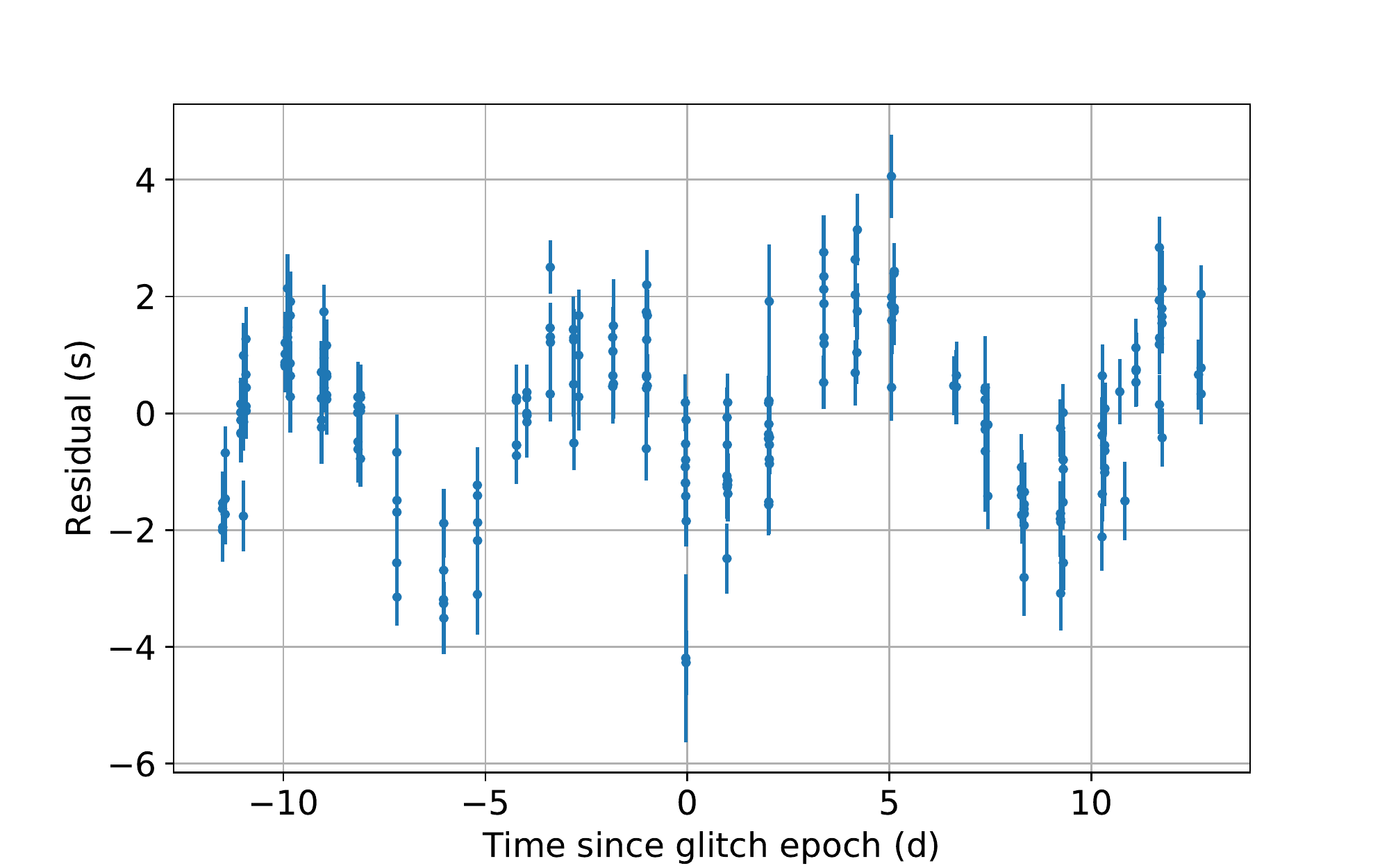}
\includegraphics[width=3.0in]{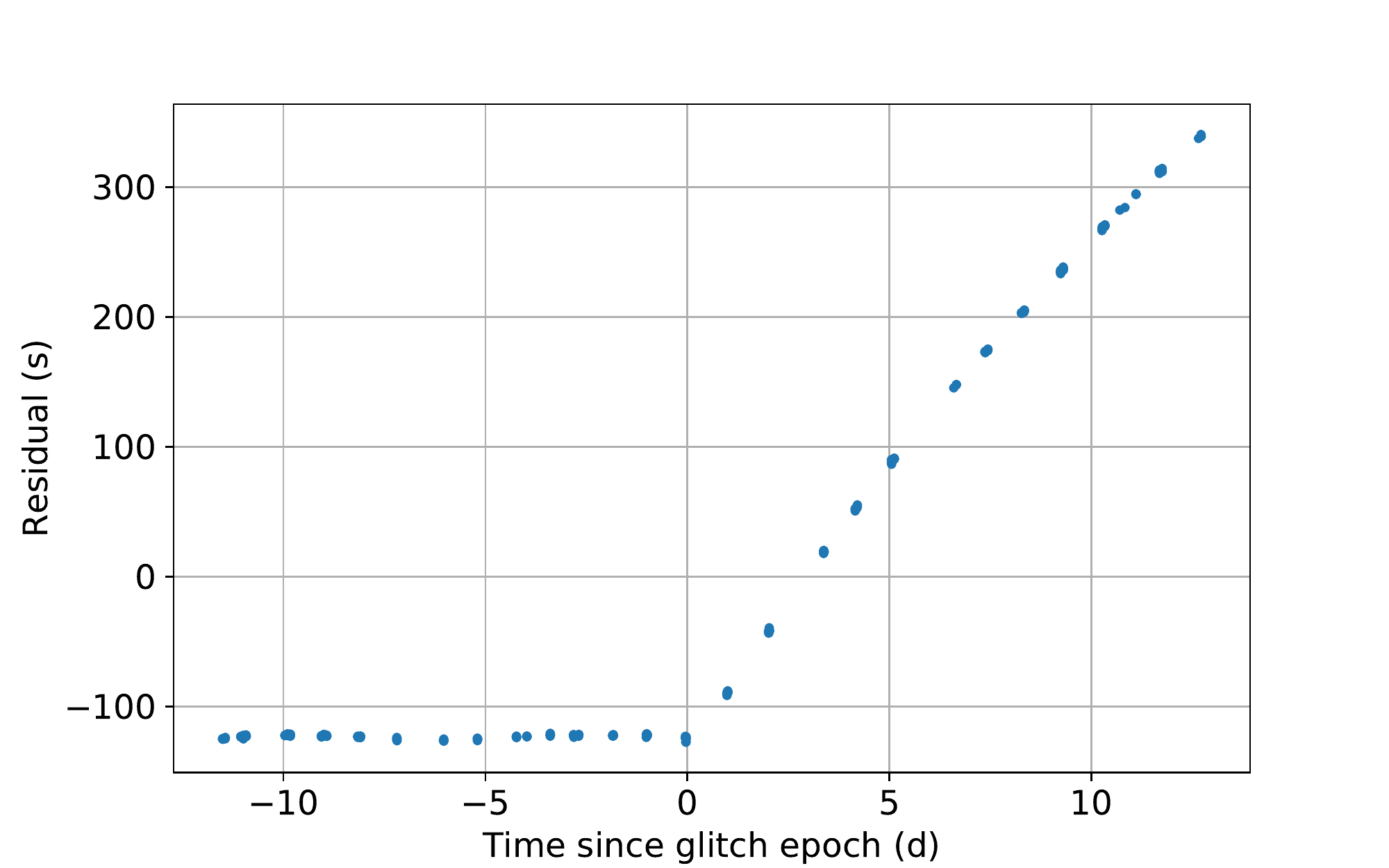}
\caption{Coherent timing of glitch 2 (segment B). Left: Residuals to a timing model including the glitch. Substantial red noise is visible in the residuals. Right: Residuals to the same model with $\Delta\nu$, $\Delta\dot{\nu}$ set to 0 to show the effect of the glitch. \label{fig:resids2}}
\end{figure}

\begin{figure}
\centering\includegraphics[width=6.5in]{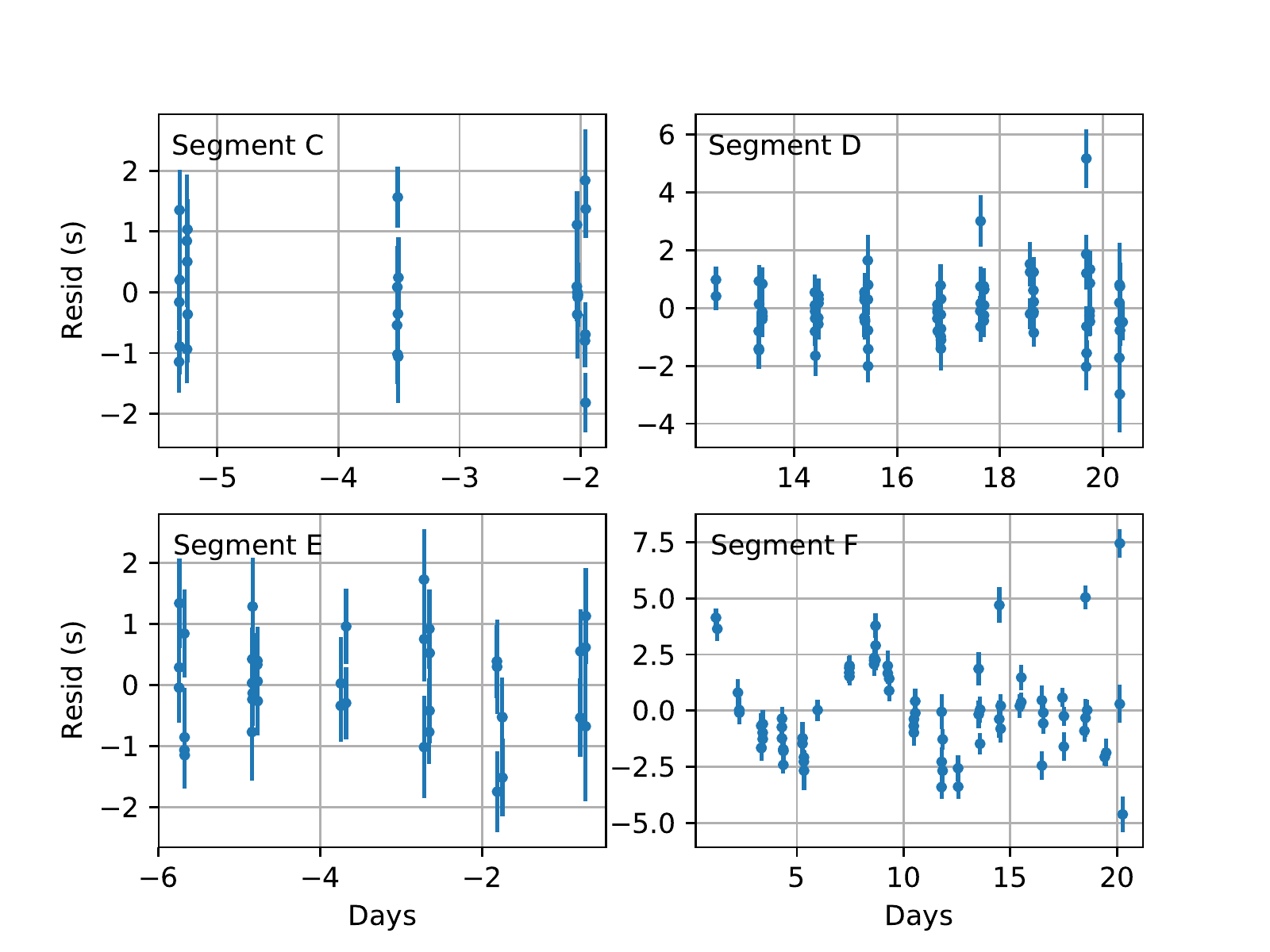}
\caption{Residuals to the coherent timing models for segments C, D, E, and F.\label{fig:residsCDEF} 
}
\end{figure}

\begin{figure}
\centering
\includegraphics[width=6.5in]{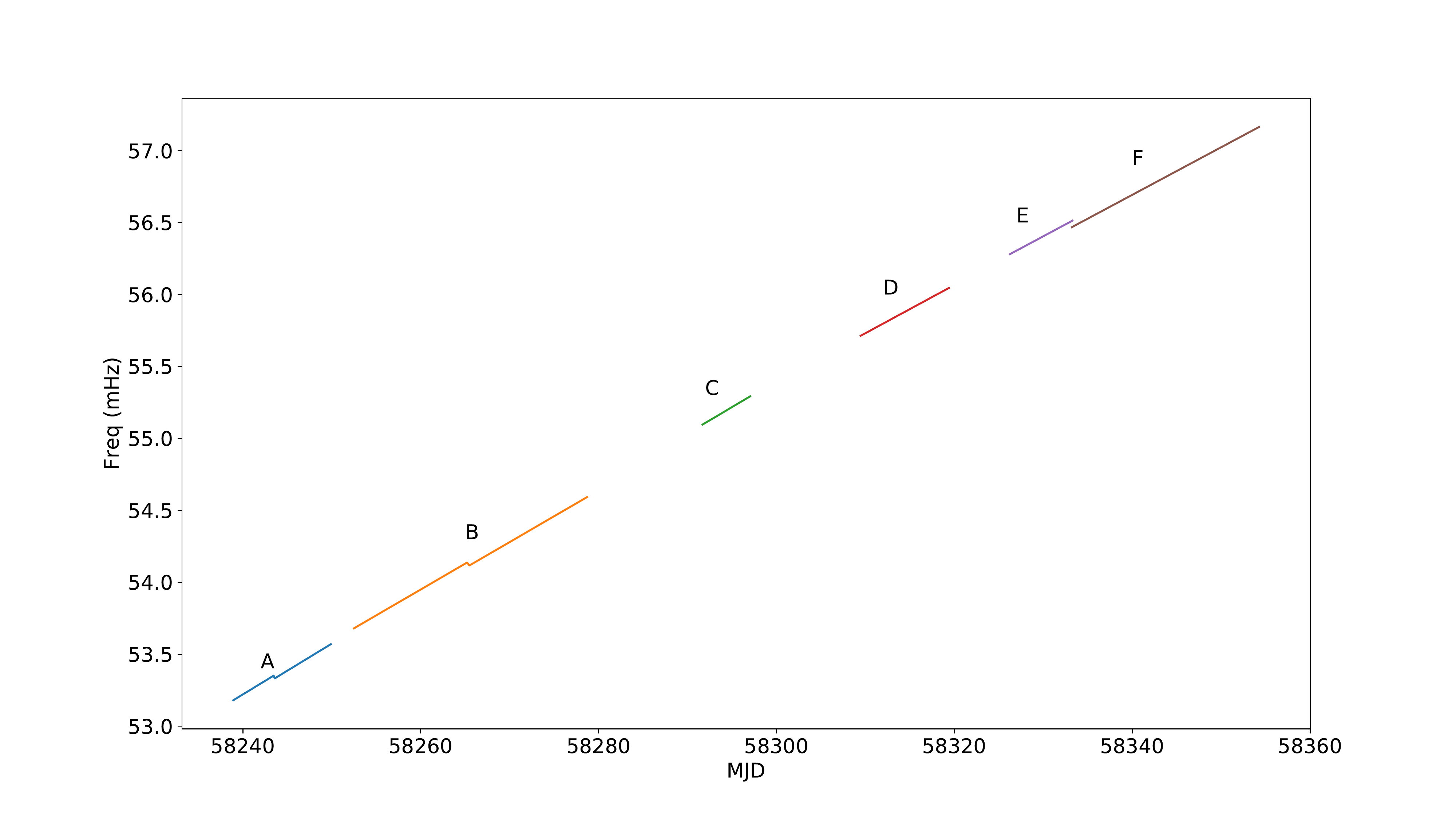}
\caption{Piecewise coherent spin-frequency model for \ngc{}.  Each color represents a segment of data with a coherent timing model fit (denoted by the letters).  \label{fig:coherentmodel}}
\end{figure}

Because of the significant red noise, the rapid spin evolution, and the gaps between observations, generating a phase-connected timing model for the full dataset is not possible.  However, we were able to make phase-connected timing models for several contiguous sections of data (which we label as segments A--F), including spans that cover the first (segment A) and second (segment B) glitches.  
For later segments (C--F) of observations, we fitted simple models with $\nu$ and $\dot{\nu}$. We use \textsc{Tempo2} \citep{TEMPO2} to fit timing models to the TOAs from each segment. The timing model parameters for each segment are presented in Tables \ref{tab:glitch1}, \ref{tab:glitch2}, and \ref{tab:glitch3} and the modeled spin-frequency evolution is displayed in Figure \ref{fig:coherentmodel}. The TOA residuals for each model are presented in Figures \ref{fig:resids}, \ref{fig:resids2}, and \ref{fig:residsCDEF}.

We considered the possibility of a glitch between segments C and D, but rejected that hypothesis since the extrapolated fits don't appear to require one. For segments E and F, we were able to get a good measurement of the frequency step at the glitch.  

The coherent timing models shown in Figures \ref{fig:resids}, \ref{fig:resids2}, and \ref{fig:residsCDEF} clearly demonstrate that there are sudden pulse frequency changes. However, it is conceivable that such features could be just a manifestation of the red noise in the system.To test this, we simulated red noise residuals. Most accretion-powered pulsars show white torque noise \citep{Bildsten}, which gives a 
pulse phase power spectrum with an index of $-4$, while some are even steeper. We made multiple realizations of residuals with red noise spectral indices of $-3$, $-4$, and $-5$, with normalizations set to match the RMS of the observed residuals. The simulations with index $-3$ sometimes showed cusp-like features, but never with the large difference in scale between the red noise wiggles and the glitch residuals that we see in Figures 7 and 8. For more realistic red noise spectral indices of $-4$ and $-5$, the red noise is much smoother and extremely unlikely to be mistaken for a glitch. We therefore conclude that the timing features we observe are best characterized as glitches.



\begin{table}
\caption{Parameters for Glitch 1 (segment A) \label{tab:glitch1}}
\begin{center}
\begin{tabular}{ll}
\hline\hline
\multicolumn{2}{c}{Fit and data-set} \\
\hline
MJD range\dotfill & 58238.9--58249.9 \\ 
Data span (d)\dotfill & 10 \\ 
Number of TOAs\dotfill & 385 \\
RMS timing residual (s)\dotfill & 0.607 \\
Reduced $\chi^2$ value \dotfill & 1.6 \\
\hline
\multicolumn{2}{c}{Fixed Quantities} \\ 
\hline
Right ascension, $\alpha$ (hh:mm:ss, J2000)\dotfill & 00:55:04.86 \\ 
Declination, $\delta$ (dd:mm:ss, J2000)\dotfill & $-$37:41:43.7 \\ 
Frequency second derivative, $\ddot{\nu}$ (Hz s$^{-2}$)\dotfill & 0 \\ 
Epoch of frequency determination (MJD)\dotfill & 58243.5 \\ 
Glitch epoch (MJD)\dotfill & 58243.509876 \\ 
\hline
\multicolumn{2}{c}{Measured Quantities} \\ 
\hline
Pulse frequency, $\nu$ (mHz)\dotfill & 53.35285(19) \\ 
Frequency first derivative, $\dot{\nu}$ (Hz s$^{-1}$)\dotfill & 4.358(13)$\times 10^{-10}$ \\ 
Glitch $\Delta\nu$\dotfill & $-$2.340(20)$\times 10^{-5}$ \\ 
Glitch $\Delta\dot{\nu}$\dotfill & 1.41(13)$\times 10^{-11}$ \\ 
\hline
\multicolumn{2}{c}{Derived Quantities} \\ 
\hline
$\Delta\nu/\nu$\dotfill & $-4.39(4) \times 10^{-4}$\\
$\Delta\dot{\nu}/\dot{\nu}$\dotfill & $3.2(3) \times 10^{-2}$\\
\hline
\end{tabular}
\end{center}
\end{table}

\begin{table}
\caption{Parameters for Glitch 2 (segment B) \label{tab:glitch2}}
\begin{center}
\begin{tabular}{ll}
\hline\hline
\multicolumn{2}{c}{Fit and data-set} \\
\hline
MJD range\dotfill & 58252.5--58278.7 \\ 
Data span (d)\dotfill & 26 \\ 
Number of TOAs\dotfill & 235 \\
RMS timing residual ($\mu s$)\dotfill & 1.36 \\
Reduced $\chi^2$ value \dotfill & 7.0 \\
\hline
\multicolumn{2}{c}{Fixed Quantities} \\ 
\hline
Right ascension, $\alpha$ (hh:mm:ss, J2000)\dotfill & 00:55:04.86 \\ 
Declination, $\delta$ (dd:mm:ss, J2000)\dotfill & $-$37:41:43.7 \\ 
Frequency second derivative, $\ddot{\nu}$ (Hz s$^{-2}$)\dotfill & 0 \\ 
Epoch of frequency determination (MJD)\dotfill & 58265 \\ 
Glitch epoch (MJD)\dotfill & 58265.25263 \\ 
\hline
\multicolumn{2}{c}{Measured Quantities} \\ 
\hline
Pulse frequency, $\nu$ (mHz)\dotfill & 54.12969(9) \\ 
First derivative of pulse frequency, $\dot{\nu}$ (s$^{-2}$)\dotfill & 4.1603(17)$\times 10^{-10}$ \\ 
Glitch $\Delta\nu$\dotfill & $-$2.990(14)$\times 10^{-5}$ \\ 
Glitch $\Delta\dot{\nu}$\dotfill & 1.24(3)$\times 10^{-11}$ \\ 
\hline
\multicolumn{2}{c}{Derived Quantities} \\
\hline
$\Delta\nu/\nu$\dotfill & $-5.52(2) \times 10^{-4}$\\
$\Delta\dot{\nu}/\dot{\nu}$\dotfill & $2.98(6) \times 10^{-2}$\\
\hline
\end{tabular}
\end{center}
\end{table}

\begin{table}
\caption{Coherent timing model parameters for segments C--F\label{tab:glitch3}}
\begin{center}
\begin{tabular}{lllll}
\hline\hline
\multicolumn{5}{c}{Fit and data-set} \\
\hline
Sequence              & C & D & E & F \\
MJD range\dotfill     & 58292.6--58296.0 & 58310.4--58318.4 & 58327.2--58332.3 & 58334.2--58353.2 \\ 
Data span (d)\dotfill & 3.4              & 8               & 5.1               & 19 \\ 
Number of TOAs\dotfill & 28              &  89             & 39                & 76 \\
RMS  resid. (s)\dotfill & 0.93           &  0.88           & 0.73              & 2.0 \\
Reduced $\chi^2$  \dotfill & 3.0         &  2.3            & 1.5               & 15.5 \\
\hline
\multicolumn{5}{c}{Fixed Quantities} \\ 
\hline
R.A., $\alpha$ \dotfill & \multicolumn{4}{c}{00:55:04.86} \\ 
Decl., $\delta$ \dotfill & \multicolumn{4}{c}{$-$37:41:43.7} \\ 
$\ddot{\nu}$ (s$^{-3}$)\dotfill & \multicolumn{4}{c}{0} \\ 
Epoch (MJD)\dotfill   & 58298             & 58298           & 58333            & 58333 \\ 
\hline
\multicolumn{5}{c}{Measured Quantities} \\ 
\hline
$\nu$ (mHz)\dotfill & 55.3280(8)  & 55.3295(5)   & 56.5046(3)       & 56.46146(13) \\ 
$\dot{\nu}$ (Hz s$^{-1}$)\dotfill & 4.25(3)$\times 10^{-10}$ & 3.877(3)$\times 10^{-10}$ & 3.838(9)$\times 10^{-10}$ & 3.8299(14)$\times 10^{-10}$\\ 
\hline
\multicolumn{5}{c}{Derived Quantities} \\
\hline
$\Delta\nu$ (Hz)\dotfill & \multicolumn{2}{c}{(D-C) $1.5(9) \times 10^{-6}$} & \multicolumn{2}{c}{(F-E) $-4.32(3) \times 10^{-5}$} \\
$\Delta\nu/\nu$\dotfill &  \multicolumn{2}{c}{$2.7(16) \times 10^{-5}$} &  \multicolumn{2}{c}{$-7.66(6) \times 10^{-4}$} \\
\hline
\end{tabular}
\end{center}
\end{table}

\section{Spectral characteristics}
\label{sec:spec}

Spectral analysis was performed with {\tt XSPEC} version 12.10 \citep{arnaud96}, and with the \nicer\ response files version 0.06.  For these analyses, the data filtering is more stringent than that of the timing analyses presented above. This is in an effort to minimize non-astrophysical background that affects the spectra.  First, we limit the energy range to 0.4--10 keV because of significant optical loading affecting channels below 0.4 keV, especially when \nicer{} is observing at low Sun angles.  In addition to the standard filtering described in Section~\ref{sec:obs}, we also exclude the detectors with \emph{DET\_ID} 14, 34, and 54, which often suffer from optical loading. Finally, we exclude periods with the conditions: ${\tt COR\_SAX}<1.5$, ${\tt FPM\_OVERONLY\_COUNT}>1.0$, and ${\tt FPM\_OVERONLY\_COUNT}>\left(1.52\times{\tt COR\_SAX}^{-0.633}\right)$. This last condition is an empirical relation obtained from a large number of ``blank sky'' pointings to characterize \nicer\ background over a wide range of orbital environments. Together, these remove time intervals with high particle background.

We generated a background spectrum specific to small time intervals using a library of spectra populated with \nicer{} observations of 7 blank-sky regions, corresponding to the \textit{RXTE} Background regions \citep{Jahoda06}. The background spectra we obtain are a combination of these blank-sky spectra weighted according to space-weather observing conditions common to both the \ngc\ and background-field observations.  

To each ObsID spectrum with its corresponding background, we fit a simple absorbed power law to study the spectral evolution over the several weeks of \nicer{} observations of \ngc{}. No obvious spectral changes can be determined at the epochs of the glitches. However, large variations of the background count rates estimated for each ObsID may hamper the precise study of the evolution.

Instead we focus our study on the spectral evolution of the pulsed component only. For each ObsID, we extract spectra from the on-pulse phase range (0.75--1.25) and off-pulse phase range (0.25--0.75; see Figure~\ref{fig:profiles}). The off-pulse spectra, used as backgrounds for the on-pulse spectra, therefore permit subtraction of the non-astrophysical background and of the other X-ray sources within the \nicer{} field of view. The typical flux of the pulsed component is $f_{\rm X}\sim 2 \times 10^{-12}$ erg s$^{-1}$ cm$^{-2}$ (0.4--10 keV).

\begin{figure}
\centering
\includegraphics[width=4.0in]{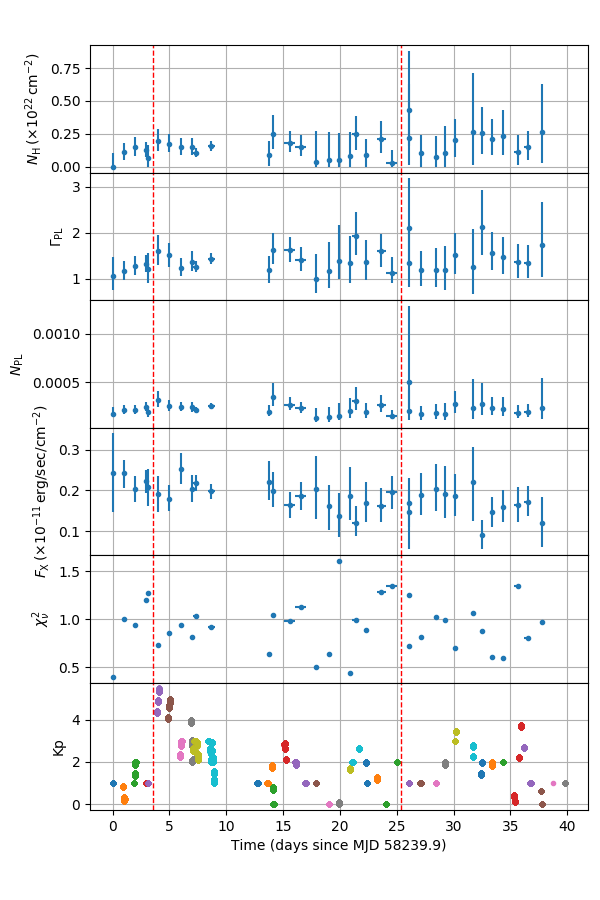}
\caption{Pulsed component spectral evolution spanning segments A and B, over glitches 1 and 2. Each spectrum was fitted with an absorbed power law of index $\Gamma_{\rm PL}$ and normalization $N_{\rm PL}$. Vertical error bars are 90\% confidence level uncertainties on the parameters, while the horizontal error bars represent the span of the ObsIDs. The dashed vertical lines show the epochs of glitches 1 and 2. The bottom panel shows the Kp space-weather index, an indicator of high backgrounds in \nicer{} data. \label{fig:spec}}
\end{figure}

We fit each of these On-minus-Off spectra with a simple absorbed power law. The best fit parameters are shown in Figure~\ref{fig:spec}, together with the reduced $\chi^{2}$ values. No significant spectral change is observed at the time of glitches 1 and 2. The observations around glitch 3 were shorter and provided low signal-to-noise spectra for a similar study.  We therefore conclude that no detectable radiative changes were associated with the glitches.  

\section{Discussion}
\label{sec:discussion}

\subsection{Characterization of the glitches}

We measured three glitches interrupting the rapid spin up of \ngc, with magnitudes 
$\Delta\nu = -23, -30, \,\mathrm{and} -43 \; \mu$Hz. In the first two glitches, our best-fit 
coherent timing models include frequency derivative steps of $\Delta\dot{\nu} = 1.4 \; \mathrm{and} \; 1.2 \times 10^{-11}$ Hz s$^{-1}$, while in the third (more poorly sampled) glitch, we observe no significant $\Delta\dot{\nu}$.

With approximately daily sampling, we do not catch any of the glitches as they happen.  We can generally
only constrain the epoch of the glitch to about $\pm 0.5$ d accuracy, unless we impose the constraint that
the pulse phase be continuous through the glitch, which gives a couple of more precise possible glitch epochs.
{There are significant unmodeled residuals (red noise) evident in the coherent fits to the glitches 2 and 3; however they are much smaller than the inferred glitch-induced component.} We fit for a $\Delta\dot{\nu}$ at the glitch, but the nature of the red noise could include glitch recovery, accretion torque noise, and Doppler shifts from the orbit of the neutron star. We are unable to distinguish among these possibilities at this point. We note that the residuals look periodic in Figure \ref{fig:resids2}.  If fitted with an orbit, one gets a period of $P_B = 7.3$ d, with a semimajor axis $(a_1 \sin i)/c = 1.8$ lt-s.  This does not seem like a plausible orbit for the neutron star as it implies a median companion mass of only 0.07 $M_\sun$ (for an inclination of 60$^\circ$), when the companion mass is known to be 20 $M_\sun$ or larger. In addition, this oscillation does not fit the other datasets with the same period and phase, so we conclude that it is not due to an orbit.

\subsection{Nature of the glitches}

\begin{figure}
\centering
\includegraphics[width=0.9\textwidth]{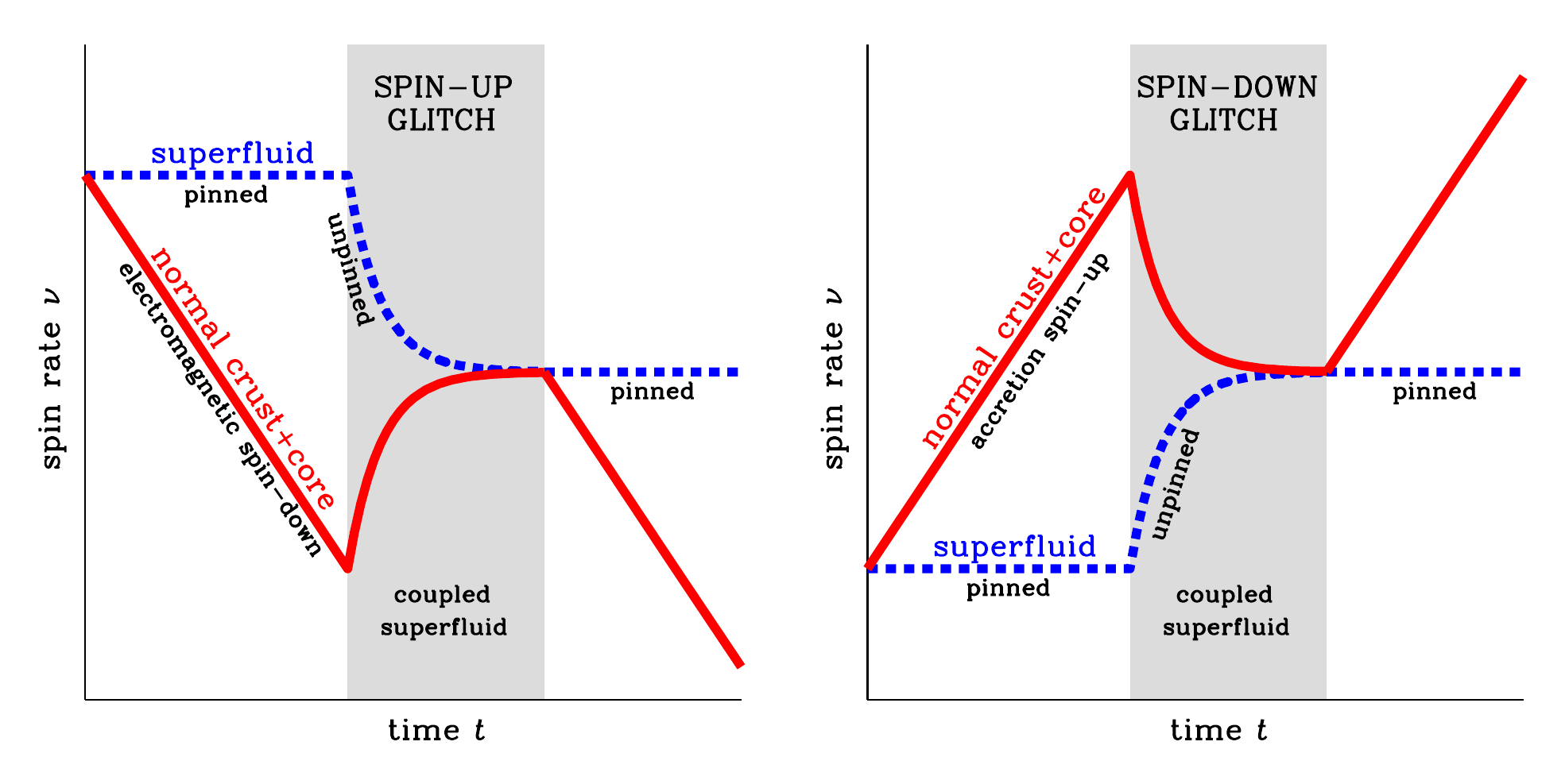}
\caption{Left: Schematic of spin-up glitch. Pulsar spin frequency (solid line) is observed to decrease due to energy loss from electromagnetic radiation, while a superfluid component within the star is pinned and does not spin down with the rest of the star (dashed line). When the superfluid component unpins and transfers angular momentum to the rest of the star, a spin-up glitch occurs. The pulsar then continues to spin down when the superfluid becomes pinned again. Right: Schematic of spin-down glitch.  Analogous to a spin-up glitch, except in this case the pulsar spin frequency (solid line) is observed to increase due to accretion torque and a spin-down glitch occurs when the superfluid gains angular momentum (dashed line). \label{fig:glitch_schematic}}
\end{figure}

The glitches measured in \ngc{} are unique amongst
glitches observed to date. 
The lack of strong evidence for any radiative change associated with the glitch in \ngc{} suggests a glitch mechanism internal to the neutron star rather than a magnetospheric mechanism. Even in the case of magnetar glitches, where only some glitches are accompanied by radiative changes, they are thought to be caused by a source within the neutron star \citep{2014ApJ...784...37D}. The internal mechanism that produces glitches in non-accreting pulsars is thought to be due to a superfluid in the pulsar's inner crust and possibly core. The free neutrons present there are expected to be in a superfluid state because the typical crust temperature is well below the critical temperature for superfluidity. Unlike normal matter, superfluid matter rotates by forming vortices whose areal density determines the spin rate of the superfluid.  These vortices are normally pinned to the lattice of crust nuclei, and as a result, while the rest of the star slows down owing to electromagnetic dipole radiation, the superfluid does not. A lag develops between the stellar spin rate and that of the superfluid. When this lag reaches a critical value, superfluid vortices unpin and transfer angular momentum to the rest of the star, causing the stellar spin rate to increase, i.e., a spin-up glitch (\citealt{1975Natur.256...25A}; see, e.g., \citealt{2015IJMPD..2430008H}, for a review; see left panel of Fig.~\ref{fig:glitch_schematic}).

The above mechanism can still apply to the case of a spin-down glitch (see, e.g., \citealt{1980PThPS..69..376P}; see right panel of Fig.~\ref{fig:glitch_schematic}), such as that seen here for \ngc{}. Here the pulsar spin frequency increases with time due to accretion (pre-glitch $\dot{\nu}=4\times 10^{-10}\mbox{ Hz s$^{-1}$}$), with a characteristic spin-up timescale $\tau_{\rm c}\equiv\nu/2\dot{\nu}=2\mbox{ yr}$. Because of this short timescale, we might expect a lag to build up quickly, where in our case most of the star spins faster than the superfluid. Once the lag is too great, the superfluid receives angular momentum from the rest of the star, and the star undergoes a spin-down glitch. This can be shown from the calculation of, e.g., \citet{2012PhRvL.109x1103A}, by simply replacing the torque due to dipole radiation with that due to accretion. Such an interpretation provides a constraint on the superfluid fraction in the neutron star \citep{1999PhRvL..83.3362L,2012PhRvL.109x1103A,2013PhRvL.110a1101C} and even potentially a determination of the neutron star mass \citep{2015SciA....1E0578H,2017nuco.confa0805H}. The former yields $I_{\rm sf}/I\approx (m^{\rm eff}_{\rm n}/m_{\rm n})(-\Sigma_i\Delta\nu^i/\nu)
(2\tau_{\rm c}/t_{\rm obs})\approx 0.02\,(m^{\rm eff}_{\rm n}/m_{\rm n})(\mbox{116 d}/t_{\rm obs})$, where the superfluid and total moments of inertia are $I_{\rm sf}$ and $I$, respectively, the average effective neutron mass due to entrainment is  $m^{\rm eff}_{\rm n}/m_{\rm n}$ ($\sim 4.3$; \citealt{2012PhRvC..85c5801C}), the summation is over each glitch, and the observation time is $t_{\rm obs}$. The large glitch amplitudes ($|\Delta\nu|\approx 20-40\mbox{ $\mu$Hz}$) and frequency derivative ($\dot{\nu}=4\times 10^{-10}\mbox{ Hz s$^{-1}$}$) of \ngc{} are comparable to those of the frequent large glitching pulsars Vela and PSR~J0537$-$6910 and $\Delta\dot{\nu}\approx 10^{-11}\mbox{ Hz s$^{-1}$}$ are among the largest measured \citep{2011MNRAS.414.1679E,2018MNRAS.473.1644A,2018ApJ...852..123F}. This could suggest that \ngc{} has a similarly high glitch activity 
\citep{2011MNRAS.414.1679E,2013MNRAS.429..688Y,2017A&A...608A.131F} if the underlying glitch mechanism is the same.

 The rise time of a glitch is uncertain, due to unknown properties of superfluidity, but is thought to be $>8\,P$ \citep{2010MNRAS.405.1061S,2017MNRAS.464.4641S}. Such a limit is in accord with glitches seen in the Vela pulsar \citep{2018Natur.556..219P} and suggests a short rise time of $>150\mbox{ s}$ for \ngc{}, well within the $\sim 12$~hr of our nearest post-glitch TOA (see Section~\ref{sec:timing}). 

\subsection{Glitch recurrence timescale}

We find three glitches in 116 days of monitoring, giving a rate of one glitch per 39 days. Alternatively, the two measured interglitch times are 22 and 68 days.
\citet{2015A&A...578A..52D} consider a model of the evolution of superfluid vortices that can produce a spin-down glitch. For this system, depending on the vortex pinning force, their model predicts a maximum amplitude $2-20\,\mbox{$\mu$Hz}$, glitch recovery timescale of $10^2-10^3$~s, and recurrence time of 40~d, although the observed {spin-down} $\Delta\dot{\nu}/\dot{\nu}$ seem to be much smaller than {that estimated from this model, with the discrepancy possibly due to model extrapolations and uncertainties}.


\subsection{No evidence for an orbit}

The initial motivation for these timing observations was to determine the orbit of the neutron star. To search for a short-period orbit, we used a Lomb-Scargle periodogram of the frequency residuals (Figure \ref{fig:plotfit4}).  No significant periodic Doppler shifts are detected. 

If we assume the pulsar is a neutron star in orbit with a 30 $M_\sun$ companion (suggested masses range from 20 to 40 $M_\sun$; see \citealt{Binderetal16,Lauetal16}), for a
typical inclination of $i=60^\circ$, the mass function will be about 17.8 $M_\sun$. 
From this, we can compute the expected amplitude of the orbital Doppler shifts, given a choice of orbital period, $P_B$.  For a relatively short orbital period of a few days, the predicted
amplitude would be easily detected in our data. For example, for $P_B = 7$ d, the amplitude of the sinusoidal Doppler modulation would be 0.05 mHz, which is much larger than the frequency residuals in Figure \ref{fig:plotfit4}.  Therefore, unless the system
is very nearly face-on, short orbital periods with massive companions can be excluded. As described above, the residuals to the coherent fits may suggest a short orbit around a very low mass 
companion ($<0.1 M_\sun$), but we conclude that this is just a few cycles of a red noise process.

If, on the other hand, the orbital period is long (e.g., $\sim 400$ d, as suggested by \citealt{Lauetal16}), the amplitude of the frequency modulation would be about 0.01 mHz,
but on a longer timescale, so it would be absorbed in the $\dot{\nu}$ and $\ddot{\nu}$ 
fit to our frequency evolution.  The expected $\dot{\nu}$ for a 400 d orbit with the 
same component masses is two orders of magnitude smaller than the observed value, so
it would not contribute significantly.  But, the $\ddot{\nu}$ magnitude would be up
to $5 \times 10^{-19}$ Hz s$^{-2}$ (depending on the orbital phase and inclination), 
which is a factor of 7 below the observed $\ddot{\nu}$.  Therefore, we conclude that a 
long orbital period could be contributing to the measured $\ddot{\nu}$ but the majority 
of the observed frequency evolution is from accretion torques.

\subsection{Magnetic field and energetics}

Using a X-ray luminosity of $5\times 10^{39}\mbox{ erg s$^{-1}$}$ \citep{carpano18} and standard accretion torque of $\dot{M}r_{\rm m}^2\Omega_{\rm K}(r_{\rm m})$, where $r_{\rm m}$ is the approximate magnetosphere radius and $\Omega_{\rm K}(r_{\rm m})$ is the Kepler orbital angular frequency at $r_{\rm m}$, we estimate that the magnetic field of \ngc{} is $\sim 1\times 10^{12}\mbox{ G}$, which is in agreement with that determined from a possible detection of an electron CRSF at $E\sim 13$\,keV \citep{walton18b}, although the field can be much higher if the pulsar is near spin equilibrium, where spin-up and spin-down torques are nearly equal. In the case of a proton CRSF, the inferred field would be $2\times 10^{15}\mbox{ G}/(1+z_g)$. We note that the presence of this CRSF was called into question by \citet{koliopanos18}, demonstrating that no spectral feature was required when modeling the \textit{XMM} and \textit{NuSTAR} spectra with a multi-color thermal emission model plus a non-thermal tail.



\section{Summary}
In this article, we \replaced{reported the discovery of of}{presented evidence for} two spin-down glitches (and a tentative third one) during the spin-up of the ULX source NGC~300~ULX-1. Our \nicer{} monitoring campaign allowed us, using phase-coherent timing analyses, to determine with high confidence that the magnitudes of these two glitches are the largest among known radio pulsar glitches.  While the third glitch had an even larger magnitude ($\Delta\nu=-43\,{\rm \mu Hz}$), its identification is somewhat less secure as gaps in monitoring precluded precise determination of the glitch epoch and amplitude.  We argue that these spin-down glitches are similar in nature to the spin-up glitches observed in spinning-down radio pulsars, i.e., the glitches are caused by sudden angular momentum transfer between a superfluid component and the rest of the neutron star. This is supported by the lack of detected radiative changes, which indicates a process internal to the neutron star. Finally, despite our dense monitoring campaign, our timing solution did not permit determination of the orbit of the binary system, either because the system's inclination is near $0\deg$ or because the Doppler amplitude of a long period orbit (hundreds of days) is hidden in the frequency evolution ($\dot{\nu}$ and $\ddot{\nu}$) of the accretion torques in this system.  Continued dense monitoring would permit studying the spin period evolution and accretion mechanisms of this intriguing source, as well as detecting more spin-down glitches to characterize the superfluid components inside the neutron star.

\acknowledgments

{We thank the anonymous referee for useful comments and suggestions that improved this article}.
We thank Nishanth Anand (Thomas Jefferson High School for Science and Technology) for his work on the analysis for this paper during his SEAP summer internship at NRL. We thank Faith Huynh (MIT) for developing excellent \nicer{} quicklook plotting software used in this work. We also thank Breanna Binder for helpful discussions.  This work was supported by NASA through the \nicer{} mission and the 
Astrophysics Explorers Program.  \nicer{} work at NRL is supported by NASA. SG is supported by the Centre National d'Etudes Spatiales (CNES). WCGH is partially supported by the Science and Technology Facilities Council (STFC) in the UK.  This research has made use of data and software provided by the High Energy Astrophysics Science Archive Research Center (HEASARC), which is a service of the Astrophysics Science Division at NASA/GSFC and the High Energy Astrophysics Division of the Smithsonian Astrophysical Observatory. This research has also made use of the NASA Astrophysics Data System (ADS) and the arXiv.

%

\vspace{5mm}
\facilities{NICER}


\software{\texttt{astropy} \citep{2013A&A...558A..33A}, 
\texttt{PINT} (\url{https://github.com/nanograv/pint}),
HEAsoft (\url{https://heasarc.nasa.gov/lheasoft/}),
\textsc{Tempo2} \citep{TEMPO2}, PRESTO (\url{https://www.cv.nrao.edu/~sransom/presto/})
}

\bibliographystyle{aasjournal}
\bibliography{nicer}

\end{document}